\documentclass[showpacs,amsmath,reprint,prb,aps]{revtex4-1}

\usepackage{hyperref}
\usepackage{graphicx}

\begin{document}

\title
{Exciton diffusion, end quenching, and exciton-exciton annihilation in individual air-suspended carbon nanotubes}
\author{A.~Ishii}
\author{M.~Yoshida}
\author{Y.~K.~Kato}
\email[Corresponding author. ]{ykato@sogo.t.u-tokyo.ac.jp}
\affiliation{Institute of Engineering Innovation, 
The University of Tokyo, Tokyo 113-8656, Japan}

\begin{abstract}
Luminescence properties of carbon nanotubes are strongly affected by exciton diffusion, which plays an important role in various nonradiative decay processes. Here we perform photoluminescence microscopy on hundreds of individual air-suspended carbon nanotubes to elucidate the interplay between exciton diffusion, end quenching, and exciton-exciton annihilation processes. A model derived from random-walk theory as well as Monte Carlo simulations are utilized to analyze nanotube length dependence and excitation power dependence of emission intensity. We have obtained the values of exciton diffusion length and absorption cross section for different chiralities, and diameter-dependent photoluminescence quantum yield have been observed. The simulations have also revealed the nature of a one-dimensional coalescence process, and an analytical expression for the power dependence of emission intensity is given.
\end{abstract}

\maketitle

\section{Introduction}
Electron-hole pairs form tightly-bound excitons in single-walled carbon nanotubes\cite{Ando:1997, Spataru:2004} because of limited screening of Coulomb interaction in one-dimensional systems.\cite{Ogawa:1991} The exciton binding energy can be larger than a third of the band gap energy,\cite{Wang:2005, Maultzsch:2005, Lefebvre:2008} making them stable even at room temperature. The optical properties of carbon nanotubes are governed by these excitons, and  they exhibit many interesting phenomena such as single photon emission from quantum-dot-like states,\cite{Hogele:2008, Hofmann:2013} brightening of luminescence by trapping sites,\cite{Miyauchi:2013} and spontaneous dissociation.\cite{Kumamoto:2014} Further investigation of excitons in carbon nanotubes is important not only for fundamental understanding of their properties but also for clarifying the physics underlying such unique phenomena.

In particular, diffusion properties of excitons deserve special attention, as they determine the efficiencies of various nonradiative decay processes. They have been studied through introduction of additional quenching sites,\cite{Cognet:2007, Siitonen:2010, Yoshikawa:2010} tube length dependence of photoluminescence (PL) intensity,\cite{Moritsubo:2010, Xie:2012, Hertel:2010} spatial profile of PL images,\cite{Georgi:2009, Crochet:2012} and transient absorption microscopy,\cite{Ruzicka:2012} resulting in exciton diffusion lengths ranging from $45-240$~nm for dispersed nanotubes\cite{Cognet:2007, Siitonen:2010, Hertel:2010, Georgi:2009, Crochet:2012, Ruzicka:2012} and $140-610$~nm for air-suspended nanotubes.\cite{Yoshikawa:2010, Moritsubo:2010, Xie:2012} 

It is notable that longer diffusion lengths have been observed for air-suspended nanotubes, likely due to the pristine nature of as-grown material. With such mobile excitons, the dominant nonradiative recombination occurs at the contacts between nanotubes and the substrate, as manifested in the length dependence of PL intensities.\cite{Moritsubo:2010, Xie:2012} Such end quenching significantly reduces the emission efficiency when the diffusion length is longer than the nanotube length.

Another important nonradiative decay process mediated by exciton diffusion is exciton-exciton annihilation (EEA), which is an Auger process involving two excitons.\cite{Wang:2004prb, Ma:2005, Russo:2006} EEA strongly depends on exciton density, leading to a sublinear excitation power dependence of emission intensity.\cite{Murakami:2009prl, Siitonen:2011, Allam:2013, Xiao:2010} A number of models have been employed to describe the sublinear power dependence, such as square-law recombination-rates,\cite{Ma:2005, Wang:2006, Matsuda:2008prb1, Xiao:2010} occupation models,\cite{Murakami:2009prb2} and first-passage approaches.\cite{Srivastava:2009, Anderson:2013}

In this work, we perform systematic investigation of exciton diffusion, end quenching, and EEA in individual air-suspended carbon nanotubes by PL microscopy using an automated optical measurement system. In Sec.~\ref{sec:PL}, preparation methods for air-suspended single-walled carbon nanotubes and details of our optical system are described. Spatial scanning and chirality identification are performed on more than 3000 individual nanotubes, and statistical characteristics of PL excitation (PLE) maps and chirality distribution in our samples are discussed. Exciton diffusion and end quenching effects are examined in Sec.~\ref{sec:Diffusion}. We derive an analytic model for length dependence of PL intensity, and  exciton diffusion lengths for five chiralities are obtained by careful characterization of individual suspended nanotubes. In Sec.~\ref{sec:EEA}, EEA process is studied by performing Monte Carlo simulations and comparing with experimentally observed excitation power dependence. We find that the rate of EEA reflects the unique nature of the one-dimensional diffusion in carbon nanotubes, and an analytical expression for excitation power dependence of PL intensity is derived.

\section{Photoluminescence microscopy on air-suspended single-walled carbon nanotubes}
\label{sec:PL}
As the electronic bands of carbon nanotubes are structure-dependent,\cite{Saito} chirality identification is an imperative step in investigating their physical properties. PLE spectroscopy is a reliable method for performing such chirality assignments through determination of the absorption and emission energies.\cite{Bachilo:2002, Weisman:2003, Lefebvre:2004apa, Ohno:2006prb, Chiashi:2008} In addition, PL spectroscopy is a non-intrusive technique that can be performed on as-grown individual nanotubes without further sample processing, ideal for characterizing the pristine material.

\subsection{Preparation of air-suspended carbon nanotubes and construction of automated photoluminescence measurement system}
\begin{figure}
\includegraphics{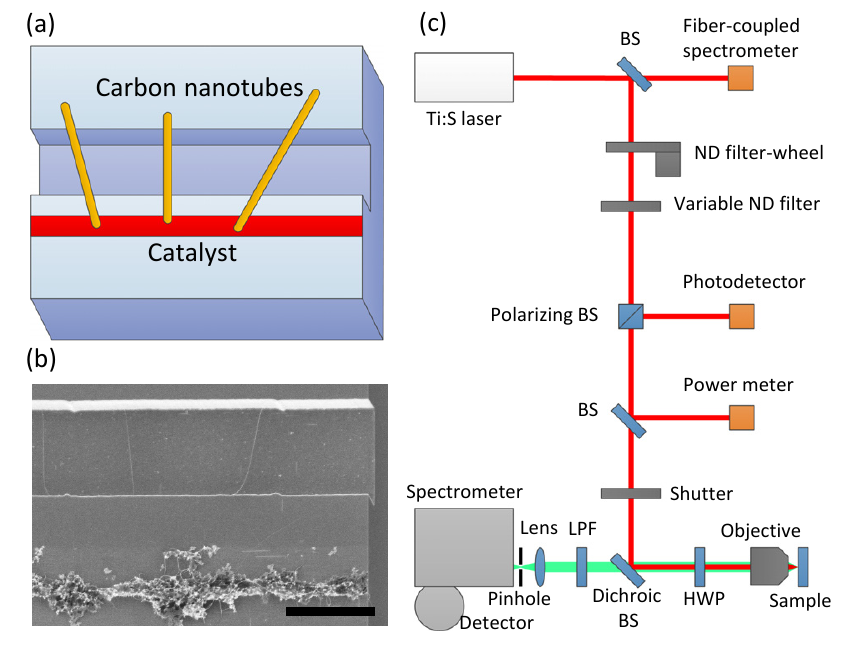}
\caption{\label{Fig1}
(Color online) (a) A schematic of a sample. 10~mm $\times$ 10~mm chips with 144~trenches with a length of 900~$\mu$m are used. (b) A scanning electron micrograph of a typical sample. The scale bar is 2~$\mu$m. (c) A schematic of the optical setup. The thin  line represents the excitation beam and the thick line indicates the PL collection path.
}
\end{figure}

\begin{figure*}
\includegraphics{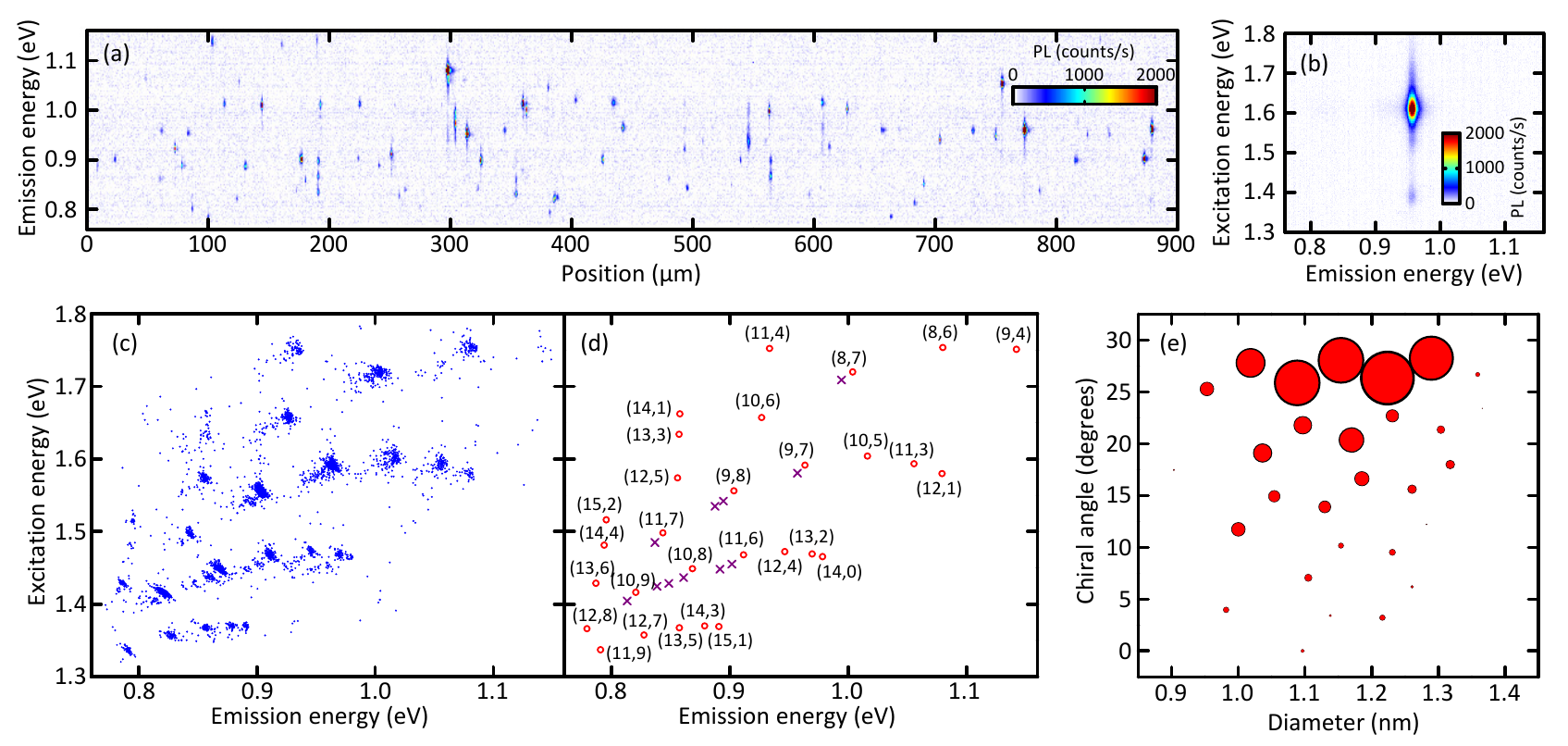}
\caption{\label{Fig2}
(a) A typical result of a trench scan with $P = 50$~$\mu$W and $E_{\mathrm{exc}} = 1.59$~eV. (b) A typical PLE map of a (9,7) nanotube with $P = 1.5$~$\mu$W. (c) PLE peak positions of 3736 individual nanotubes. (d) Averaged peak positions for each chirality. Open circles represent main spots, and cross marks indicate satellite spots. (e) Chirality distribution as a function of tube diameter and chiral angle. The area of the circles represents the population.}
\end{figure*}

Our air-suspended single-walled carbon nanotubes are grown over trenches on bare Si substrates. We perform electron beam lithography and dry etching to form the trenches as well as sample alignment marks, where the widths of the trenches range from 0.2 to 3.6~$\mu$m. Catalyst areas are patterned by another electron beam lithography step as shown in Fig.~\ref{Fig1}(a). We use 20~mg of Fe(III) acetylacetonate and 100~mg of fumed silica dispersed in 40~g of ethanol as a catalyst for nanotube growth. The catalyst particles are deposited using spin-coating and lift-off techniques, and the samples are heated in air at 400$^\circ$C for 5~minutes. 

Carbon nanotubes are synthesized by alcohol chemical vapor deposition.\cite{Maruyama:2002} Samples are brought into a quartz tube with an inner diameter of 25~mm, and its internal pressure is kept at 110~kPa with a gas mixture of 97\% Ar and 3\% H$_2$ flowing at 300~sccm. The temperature is ramped to 800$^\circ$C over 15 minutes by an electric furnace, and the gas flow and the temperature are kept constant for another 15 minutes to reduce the catalyst metal. Finally, ethanol vapor is delivered into the quartz tube for 10 minutes by switching the gas flow to bubble through a bottle of ethanol.\cite{Huang:2004} Figure~\ref{Fig1}(b) is a scanning electron micrograph of a sample after the nanotube growth process.

Because we need to characterize a large number of nanotubes, an automated confocal PL measurement system [Fig.~\ref{Fig1}(c)] has been constructed. A continuous-wave Ti:sapphire (Ti:S) laser is used for excitation, whose wavelength is controlled by a motorized linear actuator attached to the birefringent filter of the laser. The output beam is split into two paths, and the transmitted beam enters a fiber-coupled miniature spectrometer for monitoring the laser wavelength. The reflected beam passes through a neutral density (ND) filter mounted on a motorized filter-wheel which allows the excitation power $P$ to be tuned over six orders of magnitude. To fine-tune the power, we use a continuously variable ND filter placed on a translation stage with a motorized linear actuator. A polarizing beam splitter (BS) cleans the polarization and allows collection of reflected beam from the sample. 

The beam is split into two paths again, and one of the beams enters a power meter. Calibration has been performed so that the actual excitation power on the samples can be obtained. With a feedback control using the variable ND filter and the power meter, the excitation power can be tuned within an error of less than 1\%. Passing through a shutter used for background subtraction, the transmitted beam is directed towards the sample by a dichroic BS which has a cut-on wavelength at 980~nm, and then its polarization is rotated by a half-wave plate (HWP) mounted on a motorized rotation stage. We use an objective lens with a numerical aperture of 0.8 and a focal length of 1.8~mm to focus the excitation beam onto the surface of the samples with a spot size of  $\sim$1~$\mu$m, and the same lens is also used to collect emission from the nanotubes. The wavelength-dependent $1/e^2$ diameter of the focused laser has been characterized by performing PL line scans perpendicular to a suspended nanotube.

The samples are mounted on an automated three-dimensional stage, which is used for focusing as well as sample scanning. PL emitted from the nanotubes transmits the dichroic BS, and a longpass filter (LPF) with a cut-on wavelength at 950~nm is used for further laser rejection. PL from the sample is focused by a lens with a focal length of 50~mm, and it passes through a confocal pinhole with a diameter of 150~$\mu$m. We use a liquid-nitrogen-cooled 512-pixel linear InGaAs photodiode array attached to a 300-mm spectrometer with a 150~lines/mm grating blazed at 1.25~$\mu$m to obtain the PL spectra. 

The reflected beam from the sample traces back the same path as the excitation beam and is detected by a photodiode. By performing reflectivity scans, position offsets and rotation angles of the samples can be determined from the alignment marks. In addition, by bringing the laser focus at three different positions on the surface, tilt angles are obtained. Coordinate transformation can be performed from the results of these measurements, allowing the entire area of the samples to be scanned while keeping it in focus. All measurements in this work are performed at room temperature in air. The same optical system has been used for measurements on transistor devices.\cite{Kumamoto:2014, Yoshida:2014}

\subsection{Statistical characterization of individual carbon nanotubes by photoluminescence microscopy}

The suspended nanotubes are located by line scans along the trenches [Fig.~\ref{Fig2}(a)], and PLE measurements are performed for each nanotube. The PLE maps of individual nanotubes present distinct peaks in both emission and excitation energies as shown in Fig.~\ref{Fig2}(b), and $E_{11}$ and $E_{22}$ energies of the nanotubes are obtained from Lorentzian fits. As PLE maps with multiple peaks or significant broadening may come from bundled tubes\cite{Tan:2007} and nanotubes with defects, such nanotubes are excluded from further measurements. These scans are performed automatically overnight, where typically 36 trench scans are done in 9 hours while 300 PLE maps are taken in 10 hours. The positions and chiralities for thousands of nanotubes are recorded into a list, allowing for a statistical analysis.

In Fig.~\ref{Fig2}(c), peak positions in the PLE maps are plotted for all of the individual nanotubes we have measured, where high density spots corresponding to different chiralities can be seen. Chiralities are assigned by utilizing the results on ensembles of nanotubes,\cite{Ohno:2006prb, Lefebvre:2004apa, Weisman:2003} and the averaged peak positions for each chirality are summarized in Fig.~\ref{Fig2}(d) and Table~I. We note that both $E_{11}$ and $E_{22}$ energies reported earlier\cite{Ohno:2006prb} are slightly redshifted compared to our results.

Interestingly, the PLE peak distribution in Fig.~\ref{Fig2}(c) shows satellite spots with slightly lower energies. The energy shifts between the main spots and the first satellites do not depend much on chirality, and the average values are 8.1 and 12.3~meV for $E_{11}$ and $E_{22}$, respectively. For chiralities with large population such as (9,8) and (10,8), there are second and third satellite spots similar to the first satellite spots but with decreasing populations. As we do not see any apparent differences in PLE maps between the main spot and the satellites except for the slight redshifts, we speculate that these satellites come from bundles of nanotubes with the same chirality. This interpretation is consistent with the redshifted values reported in Ref.~\citenum{Ohno:2006prb}, where samples without catalyst patterning are used and more bundles are expected.

It is also interesting that the distributions of the peaks for the same chiralities are not isotropic, but show up as elongated ellipses corresponding to anticorrelated $E_{11}$ and $E_{22}$. This cannot be explained by dielectric screening effects, but seems consistent with the effects of slight strain\cite{Yang:1999, Huang:2008, Souza:2005} or bending.\cite{Koskinen:2010}

The automated scans allow us to determine the chirality distribution of semiconducting nanotubes in our samples. Trench scans are repeated three times using laser photon energies $E_{\mathrm{exc}} =$~1.46, 1.59, and 1.75~eV to locate nanotubes with different $E_{22}$ resonances. We carefully compare the results of the scans to verify that the same nanotube is only counted once, and assign chiralities by performing PLE measurements. The chirality distribution is obtained by counting the number of nanotubes, and the results are plotted in Fig.~\ref{Fig2}(e) as a function of nanotube diameter $d$ and chiral angle. It is clear that the nanotubes with larger chiral angles show larger populations, in agreement with reports of chirality-dependent growth rates.\cite{Bandow:1998, Bachilo:2003, Miyauchi:2004, Rao:2012, Liu:2013} This result is not dependent on physical parameters such as PL quantum yield\cite{Miyauchi:2004, Bachilo:2003} or Raman scattering cross section,\cite{Jorio:2005} although PL measurements can only detect semiconducting nanotubes and the spectral range is limited by the capabilities of our laser and detector. 

\begin{table}
\caption{\label{Table1}
Average $E_{11}$ and $E_{22}$ peak energies for individual air-suspended carbon nanotubes obtained from Lorentzian fits of their PLE maps. The error values are standard deviations.
}
\begin{tabular}{c r@{}c@{}l r@{}c@{}l r@{}c@{}l r@{}c@{}l}
\hline\hline
 & \multicolumn{6}{c}{$E_{11}$} & \multicolumn{6}{c}{$E_{22}$}\\
$(n,m)$ & \multicolumn{3}{c}{(nm)} & \multicolumn{3}{c}{(meV)} & \multicolumn{3}{c}{(nm)} & \multicolumn{3}{c}{(meV)}\\
\hline
(8,6)&	1148.0&	$\pm$&	4.4&	1080.0&	$\pm$&4.2&	707.0&	$\pm$&	2.4&	1753.6&	$\pm$&	6.0\\
(8,7)&	1235.1&	$\pm$&	5.0&	1003.8&	$\pm$&4.0&	720.9&	$\pm$&	2.0&	1719.9&	$\pm$&	4.7\\
(9,4)&	1085.3&	$\pm$&	3.5&	1142.3&	$\pm$&3.7&	708.0&	$\pm$&	5.6&	1751.0&	$\pm$&	13.9\\
(9,7)&	1286.8&	$\pm$&	3.9&	963.5&	$\pm$&2.9&	779.0&	$\pm$&	2.6&	1591.6&	$\pm$&	5.3\\
(9,8)&	1372.4&	$\pm$&	3.8&	903.4&	$\pm$&2.5&	796.8&	$\pm$&	2.6&	1556.1&	$\pm$&	5.1\\
(10,5)&	1219.8&	$\pm$&	4.1&	1016.4&	$\pm$&3.4&	773.0&	$\pm$&	3.3&	1604.0&	$\pm$&	6.9\\
(10,6)&	1337.7&	$\pm$&	3.7&	926.8&	$\pm$&2.6&	748.2&	$\pm$&	2.1&	1657.1&	$\pm$&	4.8\\
(10,8)&	1427.8&	$\pm$&	4.3&	868.3&	$\pm$&2.6&	855.7&	$\pm$&	2.5&	1448.8&	$\pm$&	4.2\\
(10,9)&	1511.1&	$\pm$&	5.1&	820.4&	$\pm$&2.8&	875.4&	$\pm$&	2.2&	1416.3&	$\pm$&	3.5\\
(11,3)&	1174.5&	$\pm$&	3.5&	1055.6&	$\pm$&3.1&	778.1&	$\pm$&	3.0&	1593.4&	$\pm$&	6.1\\
(11,4)&	1328.2&	$\pm$&	4.7&	933.4&	$\pm$&3.3&	707.5&	$\pm$&	2.3&	1752.4&	$\pm$&	5.7\\
(11,6)&	1360.2&	$\pm$&	3.9&	911.5&	$\pm$&2.6&	844.7&	$\pm$&	2.6&	1467.8&	$\pm$&	4.5\\
(11,7)&	1470.1&	$\pm$&	3.0&	843.3&	$\pm$&1.7&	827.6&	$\pm$&	2.0&	1498.1&	$\pm$&	3.7\\
(11,9)&	1568.0&	$\pm$&	5.5&	790.7&	$\pm$&2.8&	927.3&	$\pm$&	2.8&	1337.0&	$\pm$&	4.1\\
(12,1)&	1148.6&	$\pm$&	2.9&	1079.4&	$\pm$&2.7&	784.8&	$\pm$&	2.5&	1579.8&	$\pm$&	5.1\\
(12,4)&	1310.0&	$\pm$&	3.0&	946.4&	$\pm$&2.2&	842.1&	$\pm$&	2.1&	1472.2&	$\pm$&	3.7\\
(12,5)&	1448.7&	$\pm$&	3.6&	855.8&	$\pm$&2.2&	787.7&	$\pm$&	1.8&	1573.9&	$\pm$&	3.7\\
(12,7)&	1498.7&	$\pm$&	4.3&	827.3&	$\pm$&2.4&	913.4&	$\pm$&	1.9&	1357.3&	$\pm$&	2.8\\
(12,8)&	1591.2&	$\pm$&	3.1&	779.2&	$\pm$&1.5&	907.6&	$\pm$&	2.2&	1366.0&	$\pm$&	3.4\\
(13,2)&	1278.8&	$\pm$&	2.4&	969.5&	$\pm$&1.8&	843.9&	$\pm$&	2.1&	1469.1&	$\pm$&	3.7\\
(13,3)&	1446.6&	$\pm$&	4.5&	857.0&	$\pm$&2.6&	758.7&	$\pm$&	3.1&	1634.1&	$\pm$&	6.6\\
(13,5)&	1446.1&	$\pm$&	3.5&	857.4&	$\pm$&2.1&	906.7&	$\pm$&	1.6&	1367.4&	$\pm$&	2.4\\
(13,6)&	1576.1&	$\pm$&	3.8&	786.6&	$\pm$&1.9&	867.7&	$\pm$&	1.8&	1428.8&	$\pm$&	3.0\\
(14,0)&	1267.3&	$\pm$&	2.5&	978.3&	$\pm$&1.9&	846.1&	$\pm$&	1.6&	1465.3&	$\pm$&	2.8\\
(14,1)&	1445.5&	$\pm$&	3.8&	857.7&	$\pm$&2.2&	745.9&	$\pm$&	2.2&	1662.1&	$\pm$&	4.9\\
(14,3)&	1411.1&	$\pm$&	3.4&	878.6&	$\pm$&2.1&	905.0&	$\pm$&	1.7&	1369.9&	$\pm$&	2.6\\
(14,4)&	1561.9&	$\pm$&	3.6&	793.8&	$\pm$&1.8&	837.1&	$\pm$&	1.7&	1481.1&	$\pm$&	3.1\\
(15,1)&	1392.1&	$\pm$&	2.3&	890.6&	$\pm$&1.5&	905.7&	$\pm$&	2.1&	1369.0&	$\pm$&	3.1\\
(15,2)&	1558.6&	$\pm$&	1.8&	795.5&	$\pm$&0.9&	817.7&	$\pm$&	3.1&	1516.2&	$\pm$&	5.8\\
\hline\hline
\end{tabular}
\end{table}

\section{Exciton diffusion and end quenching}
\label{sec:Diffusion}
In air-suspended carbon nanotubes, exciton diffusion strongly influences the brightness of emission because of efficient nonradiative recombination at the nanotube ends.\cite{Moritsubo:2010, Xie:2012, Anderson:2013} The end quenching effects become particularly important when the nanotube length is shorter than or comparable to the exciton diffusion length, significantly reducing the emission efficiency. By modeling such an effect, the exciton diffusion length can be obtained from tube length dependence of PL intensity.

\subsection{Analytical model for emission intensity}
We begin by deriving an analytical expression under inhomogeneous excitation. Intrinsic decay and end quenching processes are considered to be independent, while EEA is not explicitly taken into account. For intrinsic decay, the survival probability is $S_{\mathrm{I}} (t) = \exp (-t / \tau)$, where $t$ is the time after exciton generation and $\tau$ is the intrinsic lifetime of excitons. Taking the origin of the coordinate system to be the center of the tube, the end quenching survival probability for an exciton with an initial position $z_0$ at time $t$ is given by 
\begin{widetext}
\begin{equation}
S_{\mathrm{E}} (z_0, t) = 1 - \sum_{k=0}^{\infty} (-1)^k
\left\{ \mathrm{erfc} \left[ \frac{\frac{L}{2} (1+2k) - z_0}{\sqrt{4Dt}} \right]
+ \mathrm{erfc} \left[ \frac{\frac{L}{2} (1+2k) + z_0}{\sqrt{4Dt}} \right] \right\},
\label{SurvivalEndQuenching}
\end{equation}
\end{widetext}
where $L$ is the nanotube length, $D$ is the diffusion constant of excitons, $\mathrm{erfc}(x) = 1-\mathrm{erf}(x)$ is the complimentary error function, and $\mathrm{erf} (x) = \frac{2}{\sqrt{\pi}} \int_0^x \exp (-y^2) dy$ is the error function (Appendix~\ref{app:FirstPassage}). The total survival probability for an exciton generated at $z_0$ is $S_{\mathrm{I}} (t) S_{\mathrm{E}} (z_0, t)$, and the probability that the exciton goes through intrinsic decay is
\begin{equation}
P_{\mathrm{I}} (z_0)
= \frac{1}{\tau} \int_{0}^{\infty} S_{\mathrm{I}} (t) S_{\mathrm{E}} (z_0, t) dt
= 1 - \frac{\cosh(z_0 / l)}{\cosh (L / 2l)},
\label{ProbabilityIntrinsicDecay}
\end{equation}
where $l = \sqrt{D \tau}$ is the exciton diffusion length. It is important that the initial position $z_0$ remains in Eq.~(\ref{ProbabilityIntrinsicDecay}), as it allows us to deal with inhomogeneous excitation. Using exciton generation rate profile $g(z_0)$, intrinsic decay rate integrated over the length of a nanotube is expressed as
\begin{equation}
\Gamma_{\mathrm{I}} (L)
= \int_{-L/2}^{L/2} g(z_0) P_{\mathrm{I}} (z_0) dz_0,
\label{IntrinsicDecayRate}
\end{equation}
and PL intensity from the nanotube is given by $I (L) = A \eta_0 \Gamma_{\mathrm{I}}(L)$, where $\eta_0$ is the intrinsic PL quantum yield and $A$ is a proportionality constant. We note that $A$ includes system-related factors such as collection efficiency of the objective lens, transmittance and reflectivity of the optics, and detector sensitivity. 

The generation rate profile in our system can be written as $g(z_0) = g_0 \sqrt{\frac{2}{\pi}} \frac{1}{r} \exp \left( - \frac{ 2 {z_0}^2 }{r^2} \right)$, where $g_0 = \sqrt{\frac{2}{\pi}} \frac{n_{\mathrm{c}}}{r E_{\mathrm{exc}}} \sigma_{\mathrm{ab}} P$ is the total generation rate, $r$ is the $1/e^2$ radius of the laser spot, $n_{\mathrm{c}}$ is the number of carbon atoms per unit length, $E_{\mathrm{exc}}$ is the laser photon energy, $\sigma_{\mathrm{ab}}$ is the absorption cross section per carbon atom, and $P$ is the excitation power. PL intensity is then given by
\begin{equation}
I (L)
= A \eta_0 \sigma_{\mathrm{ab}} \frac{2}{\pi} \frac{n_{\mathrm{c}}}{r^2 E_{\mathrm{exc}}} P
\int_{-L/2}^{L/2} \exp \left( - \frac{ 2 {z_0}^2 }{r^2} \right) P_{\mathrm{I}} (z_0) dz_0,
\label{PLintensity}
\end{equation}
where the unknown parameters are $l$ which determines the form of $P_{\mathrm{I}} (z_0)$ and the proportionality coefficient $A \eta_0 \sigma_{\mathrm{ab}}$. By fitting Eq.~(\ref{PLintensity}) to data, the diffusion length and the coefficient $A \eta_0 \sigma_{\mathrm{ab}}$ can be obtained. Assuming that $A$ is not strongly dependent on emission wavelength, it is possible to compare the relative values of the intrinsic PL action cross section $\eta_0 \sigma_{\mathrm{ab}}$ among different chiralities.

\subsection{Length dependence measurements and extraction of diffusion lengths}
\begin{figure}
\includegraphics{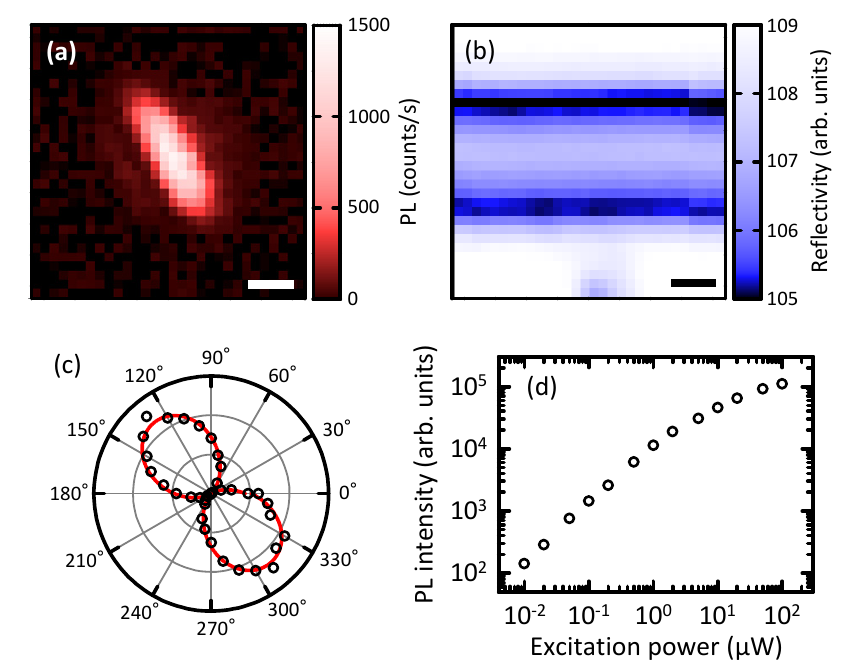}
\caption{\label{Fig3}
(Color online) (a) and (b) PL and reflectivity images, respectively, of a (9,7) nanotube. $P = 0.5$~$\mu$W and $E_{\mathrm{exc}} = 1.594$~eV are used, and the PL image is extracted at an emission energy of 0.961~eV. The scale bars are 1~$\mu$m. (c) Polarization dependence of PL intensity with $P = 0.05$~$\mu$W and $E_{\mathrm{exc}} = E_{22}$. The line is a fit. (d) Excitation power dependence of emission. PL intensity is obtained by calculating the peak area of a Loretzian fit to the emission spectrum. 
}
\end{figure}

\begin{figure}
\includegraphics{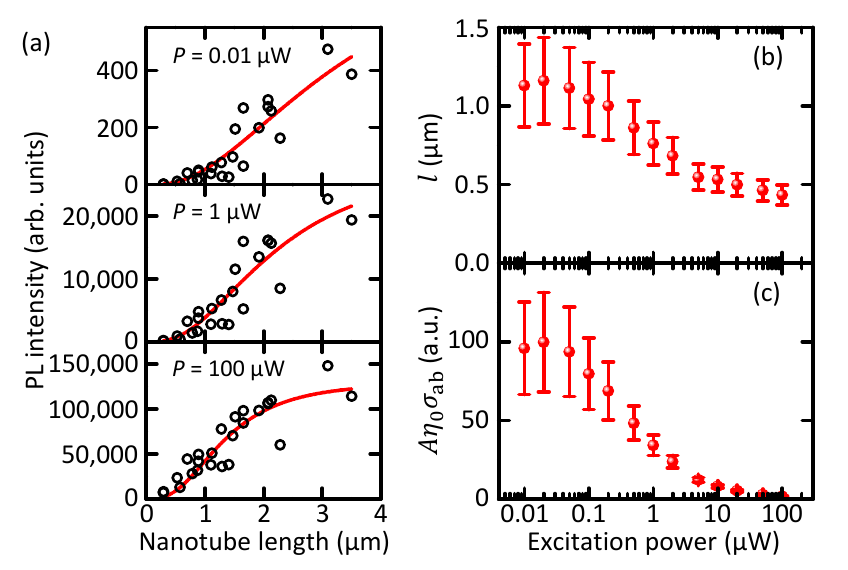}
\caption{\label{Fig4}
(Color online) (a) Length dependence of PL intensity for (9,7) nanotubes with $P =$~0.01, 1, and 100~$\mu$W. The curves are fits with Eq.~(\ref{PLintensity}). (b) and (c) Excitation power dependence of diffusion length and PL action cross section, respectively.}
\end{figure}

In order to carry out such analysis, measurements on high-quality chirality-identified tubes with various lengths are needed. Making use of the large list of nanotubes generated from the sample scans (Sec.~\ref{sec:PL}), we select bright nanotubes and perform detailed characterization by PL microscopy.

We start with PL and reflectivity imaging to confirm that the nanotubes are fully suspended over the trenches. Figure~\ref{Fig3}(a) is a typical PL image of a suspended nanotube and Fig.~\ref{Fig3}(b) is a reflectivity image which shows the trench. If a PL image of a nanotube is not centered at the trench, the nanotube is rejected from subsequent measurements because it may have fallen into the trench. For fully suspended nanotubes, high-resolution PLE maps are taken at the center of the nanotubes to determine the exact $E_{22}$ energies for excitation. The suspended length $L$ of the nanotube is obtained by measuring the polarization dependence of PL intensity\cite{Moritsubo:2010} [Fig.~\ref{Fig3}(c)]. Nanotubes with a low degree of polarization are not used in the following measurements because they may be bent or curved. After confirming that PL does not show intermittency,\cite{Matsuda:2005} PL intensity is measured as a function of power with the excitation at the $E_{22}$ energy and the polarization parallel to the tube axis [Fig.~\ref{Fig3}(d)]. The sublinear increase at high power excitations is caused by EEA, which will be extensively discussed in Sec.~\ref{sec:EEA}. Such rigorous characterization is repeated for nanotubes with various lengths and chiralities, and we have collected excitation power dependence of PL intensity for more than 400 nanotubes in total.

During the power dependence measurements, we have observed discrete blueshifts of the emission energies for powers above 100~$\mu$W. The blueshifts are chirality dependent with typical values of $10-20$~meV. The emission energy recovers reversibly upon reducing the power, suggesting that laser heating causes molecular desorption.\cite{Finnie:2005, Chiashi:2008, Xiao:2014} To avoid complications in data analysis, we keep the powers low enough to avoid such blueshifts.

We now perform the analysis on length dependence of PL intensity. In Fig.~\ref{Fig4}(a), the PL intensity of (9,7) nanotubes is plotted as a function of nanotube length for three different excitation powers of 0.01, 1, and 100~$\mu$W. At higher powers, we can see that the PL intensity starts to increase at shorter nanotube lengths, which suggests that the effective diffusion length is shorter. \cite{Moritsubo:2010, Xie:2012} This is expected as more EEA occurs at high powers, resulting in shorter exciton lifetimes.

We fit the data with Eq.~(\ref{PLintensity}) to extract the diffusion length and PL action cross section. In Figs.~\ref{Fig4}(b) and \ref{Fig4}(c), excitation power dependence of $l$ and $A \eta_0 \sigma_{\mathrm{ab}}$ are plotted, respectively. Both parameters stay nearly constant at the lowest powers, implying negligible contributions of EEA.\cite{Xie:2012} The values decrease with increasing power, which can be explained by the reduction of lifetime due to EEA. As $\eta_0 \propto \tau$ while $l \propto \sqrt{\tau}$, it is reasonable that $A \eta_0 \sigma_{\mathrm{ab}}$ drops more rapidly compared to $l$ as the power increases. We note, however, that $l$ approaches a nonzero value at the highest excitation powers, suggesting the limit of this model which does not include EEA effects explicitly.

\begin{figure}
\includegraphics{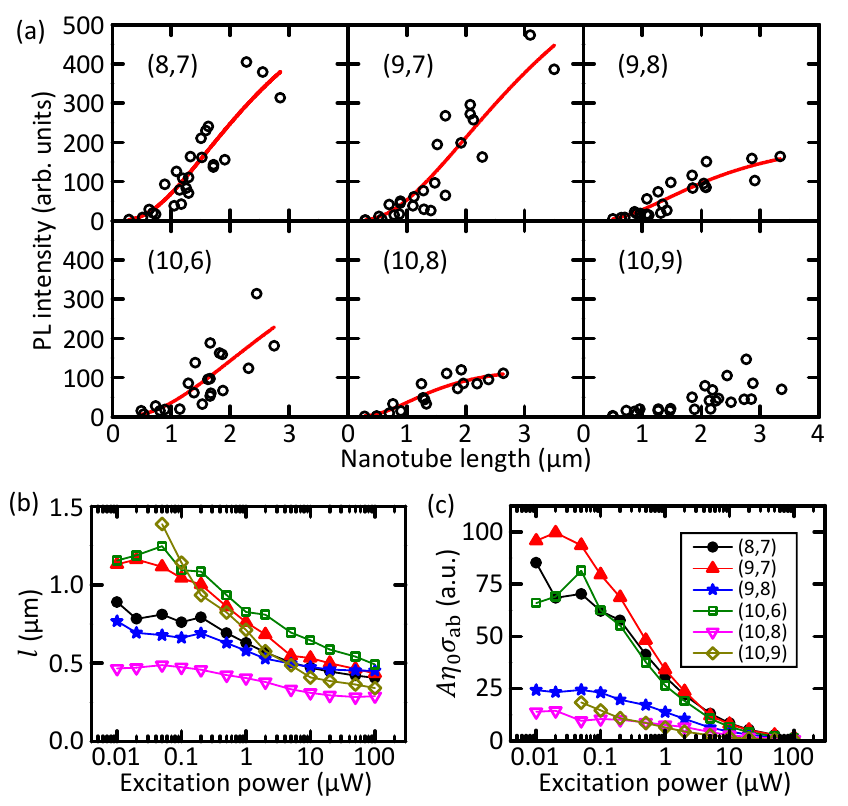}
\caption{\label{Fig5}
(Color online) (a) Length dependence of PL intensity at $P = 0.01$~$\mu$W for six different chiralities. Lines are fits, and it is not shown for (10,9) because a reliable fit has not been obtained. (b) and (c) Excitation power dependence of diffusion lengths and PL action cross sections, respectively.
}
\end{figure}

We are also able to perform such an analysis for other chiralities with large numbers of nanotubes. In Fig.~\ref{Fig5}(a), length dependence of PL intensity for chiralities (8,7), (9,7), (9,8), (10,6), (10,8), and (10,9) are shown, where the lowest excitation power is used to obtain the exciton diffusion length in the absence of EEA effects. We find that these chiralities exhibit quite different behaviors. (8,7) and (9,7) nanotubes show larger PL intensity compared to other nanotubes, while emission from (9,8) and (10,8) nanotubes saturate at shorter lengths compared to others. (10,9) nanotubes are unique, with weak PL intensity but without any sign of saturation even at the nanotube length of $\sim$3~$\mu$m. In Table~II, we summarize $l$ and $A \eta_0 \sigma_{\mathrm{ab}}$ obtained from fits by Eq.~(\ref{PLintensity}). We note that fits are unreliable for (10,9) nanotubes, likely because of diffusion length being much longer than the nanotube lengths investigated.

The fits are repeated for every excitation power, and excitation power dependence of $l$ and $A \eta_0 \sigma_{\mathrm{ab}}$ are shown in Figs.~\ref{Fig5}(b) and \ref{Fig5}(c). For all chiralities except for (10,9), we observe the low-power regions where EEA is negligible. Because of the differences of exciton effective mass, the diffusion length should depend on nanotube diameter $d$.\cite{Srivastava:2009} A clear correlation is absent, possibly because exciton diffusion is limited by disorders\cite{Siitonen:2010, Crochet:2012} resulting from air or water molecules adsorbed onto the nanotubes. Measurements in vacuum may give further insight to intrinsic exciton diffusion properties of pristine nanotubes.

\begin{table}
\caption{\label{Table2}
The values of $l$ and $A \eta_0 \sigma_{\mathrm{ab}}$ obtained from tube length dependence analysis in Sec.~\ref{sec:Diffusion}, and $\sigma_{\mathrm{ab}}$ and $A \eta_0$ given by excitation power dependence fits in Sec.~\ref{sec:EEA}.
}
\begin{tabular}{cc r@{$\pm$}l r@{$\pm$}l r@{$\pm$}l r@{$\pm$}l}
\hline\hline
 & $d$ & \multicolumn{2}{c}{$l$} & \multicolumn{2}{c}{$A \eta_0 \sigma_{\mathrm{ab}}$} & \multicolumn{2}{c}{$\sigma_{\mathrm{ab}}$} & \multicolumn{2}{c}{$A \eta_0$} \\
$(n,m)$ & (nm) & \multicolumn{2}{c}{($\mu$m)} & \multicolumn{2}{c}{(a.u.)} & \multicolumn{2}{c}{($\times 10^{-17}$~cm$^2$)} & \multicolumn{2}{c}{(a.u.)} \\
\hline
(8,7)&	1.02& 	0.89&0.20& 	85.3&24.6&	\phantom{0 }17.2&7.3&	5.1&1.7\\
(9,7)&	1.09& 	1.13&0.26& 	95.8&29.5&	\phantom{0 }17.8&9.6&	5.5&	2.0\\
(9,8)&	1.15& 	0.77&0.15& 	24.3&5.0&	\phantom{0 }7.5&3.4&		4.0&	1.4\\
(10,6)&	1.10& 	1.15&0.67& 	66.0&56.7& 	\phantom{0 }20.3&10.4&	3.9&	1.6\\
(10,8)&	1.22& 	0.47&0.12& 	13.9&2.7&	\phantom{0 }5.2&3.1&		3.0&	1.5\\
\hline\hline
\end{tabular}
\end{table}

\section{Exciton-exciton annihilation}
\label{sec:EEA}
Another important decay process governed by exciton diffusion is EEA, which causes the sublinear power dependence of PL intensity when the density of excitons is high.\cite{Wang:2004prb, Murakami:2009prl, Xiao:2010, Siitonen:2011, Allam:2013} As the model in the previous section did not explicitly take EEA into account, we compare the experimental data with Monte Carlo simulations. Now that we have obtained the diffusion lengths, the simulations can be performed without making strong assumptions or adjustable parameters.

\subsection{Comparison with Monte Carlo simulations}
In our simulations, exciton generation, diffusion, and decay processes are evaluated at short time-intervals $\Delta t$. Typically we use $\Delta t = 10^{-4} \tau$, and a smaller interval is used to increase accuracy when the exciton density is expected to be large. For every time interval, excitons are generated using $g(z_0) = g_0 \sqrt{\frac{2}{\pi}} \frac{1}{r} \exp \left( - \frac{ 2 {z_0}^2 }{r^2} \right)$ as a generation rate profile, and we let all existing excitons to diffuse with a probability given by the normal distribution $\frac{1}{\sqrt{4 \pi D \Delta t}} \exp \left( - \frac{s^2}{4 D \Delta t} \right)$ where $s$ is the displacement. We then evaluate intrinsic decay, end quenching, and EEA, which are the three types of decay processes implemented in our simulations. The intrinsic decay occurs with a probability of $\Delta t / \tau$ for all excitons, while end quenching eliminates the excitons which have diffused beyond the ends of the nanotube. EEA causes an exciton to decay when any two excitons pass by each other.

In order to perform comparison with the experiments, the simulations are executed using $L$ and $r$ obtained from the measurements while we use the values of $l$ extracted by the analysis presented in Sec.~\ref{sec:Diffusion}. The simulations run until a certain number of excitons go through intrinsic decay, and the exciton number is time-averaged after discarding the data during the initial times earlier than $\tau$. This gives the steady-state exciton number $N$, which is related to the PL emission intensity through the intrinsic decay rate $\Gamma_{\mathrm{I}} = N / \tau$.

\begin{figure}
\includegraphics{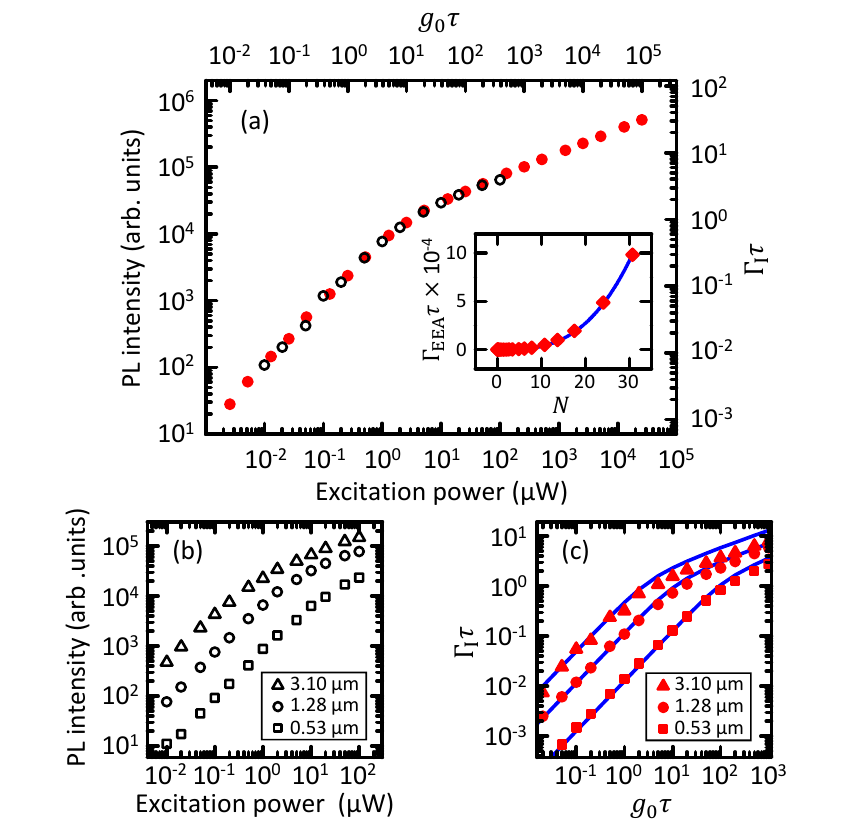}
\caption{\label{Fig6}
(Color online) (a) Excitation power dependence of PL intensity for a 0.89-$\mu$m-long (8,7) nanotube (open circles) and generation rate dependence of intrinsic decay rate in simulations (filled circles). Inset shows exciton number dependence of EEA decay rate in the same simulations. Line is a $N^3$ fit. (b) Excitation power dependence of PL intensity for (9,7) nanotubes with $L= 0.53,$ 1.28, and 3.10~$\mu$m. (c) Generation rate dependence of intrinsic decay rate for the same nanotube lengths as in (b). Lines are results of calculations by Eq.~(\ref{CubicFormula}) with the same parameters used in the simulations.
}
\end{figure}

Open circles in Fig.~\ref{Fig6}(a) show typical excitation power dependence of PL intensity in experiments. At low powers, the emission increases linearly, while at high powers EEA causes the increase to become sublinear. This power dependent behavior is reproduced in the simulations (filled circles), where the results are plotted in terms of unitless parameters $g_0 \tau$ and $\Gamma_{\mathrm{I}} \tau$. The transition to the sublinear regime can be seen as a kink in this log-log plot, and occurs at around $N = \Gamma_{\mathrm{I}} \tau = 1$ which is consistent with the PL saturation at few-exciton levels observed for pulsed excitation.\cite{ Xiao:2010}

By adjusting the ratios $P/g_0 \tau$ and $I/\Gamma_{\mathrm{I}} \tau$, we are able translate the simulation results on the plot to match the experimental data. The ratio $P/g_0 \tau$ is then related to the absorption cross section by
\begin{equation}
\sigma_{\mathrm{ab}}
= \frac{1}{\tau} \sqrt{\frac{\pi}{2}}\frac{r E_{\mathrm{exc}}}{n_{\mathrm{c}}} \left(\frac{P}{g_0 \tau}\right)^{-1},
\label{Sigma}
\end{equation}
where the only unknown parameter is $\tau$. Similarly, the ratio $I/\Gamma_{\mathrm{I}} \tau$ gives the quantum yield through
\begin{equation}
A \eta_0
= \tau \left(\frac{I}{\Gamma_{\mathrm{I}} \tau}\right).
\label{Aeta}
\end{equation}

By comparing the simulations with all the experimental data that we have, average values of $P/g_0 \tau$ and $I/\Gamma_{\mathrm{I}} \tau$ have been obtained for the five chiralities. We have computed the values of $\sigma_{\mathrm{ab}}$ and $A \eta_0$ assuming $\tau =350$ ps,\cite{Anderson:2013} and the results are listed in Table~II. The values of $\sigma_{\mathrm{ab}}$ are comparable to recent reports,\cite{Kumamoto:2014, Streit:2014} and there are large differences among these chiralities as in the case of the exciton diffusion length. Tubes with larger $d$ show smaller values of $A \eta_0$, consistent with measurements of radiative lifetimes.\cite{Miyauchi:2009} It is noted that the product of these values display good agreement with the relative values of PL action cross section $A \eta_0 \sigma_{\mathrm{ab}}$ obtained in Sec.~\ref{sec:Diffusion}, showing that the analysis is self-consistent.

We further discuss the influence of nanotube length on power dependence of PL intensities. In Fig.~\ref{Fig6}(b), data for (9,7) nanotubes with three different lengths are presented, where we observe linear increase at low powers and sublinear behavior at high powers as discussed above. Longer nanotubes show brighter emission and their kinks appear at lower excitation powers, as expected from reduced end quenching effects. The results of simulations using the lengths of these nanotubes are plotted in Fig.~\ref{Fig6}(c), and they reproduce the experimental data well. We note, however, that experimental data at higher powers tend to show brighter PL for longer tubes when compared to the simulations. This tendency may be caused by other exciton-related processes which are not taken into account in the simulations, such as EEA process involving $E_{22}$ excitons\cite{Harrah:2011b} and exciton-exciton scattering.\cite{Nguyen:2011}

\subsection{Analytical expression for power dependence of emission intensity}
Confirming that our simulations reproduce the power dependent behavior of PL intensity, we have extended the range of simulations to a higher power regime as shown in Fig.~\ref{Fig6}(a). Although such high powers are not experimentally accessible because of damage to nanotubes, the simulations can be used to illustrate the physics governing EEA in one-dimensional systems. At the extended power region, we find that the slope in the log-log plot becomes $\sim$1/3, indicating a relation $\Gamma_{\mathrm{I}} \tau = N \propto \sqrt[3]{g_0 \tau}$. This is different from the case of pulsed excitation where saturation of emission intensity is observed.\cite{Murakami:2009prl, Xiao:2010} For high power continuous-wave excitation, exciton decay is dominated by EEA and therefore the EEA rate $\Gamma_{\mathrm{EEA}} \approx g_0$ under steady-state conditions. The rate of EEA can then be described by $\Gamma_{\mathrm{EEA}} \propto N^3$, which can be confirmed by counting the number of EEA events in the simulations [inset of Fig.~\ref{Fig6}(a)].

Although treatments of EEA have sometimes involved the assumption that it is proportional to the square of exciton density as in two- and three-dimensional systems,\cite{Ma:2005, Wang:2006, Matsuda:2008prb1, Konabe:2009, Xiao:2010} the cubic dependence is characteristic of one-dimensional diffusion of excitons. This can be shown by recasting the EEA process into a model with a single particle with twice the diffusion constant and a fixed absorbing boundary.\cite{Redner, Srivastava:2009, Anderson:2013} For an initial exciton density of $n_0$, the survival probability is then given by $S_1 (t) = \mathrm{erf} \left( \frac{1/n_0}{\sqrt{8Dt}} \right)$ (Appendix~\ref{app:FirstPassage}). When $\sqrt{Dt} \gg \frac{1}{n_0}$, $S_1 (t) \approx \frac{1/n_0}{\sqrt{2 \pi D t}}$, where the $t^{-1/2}$ decay of exciton density has been observed experimentally with time-resolved measurements using pulsed excitation.\cite{Russo:2006, Allam:2013} From the time-dependence of the survival probability, the effective decay rate of exciton density $n$ becomes
\begin{equation}
\frac{dn}{dt}
= n_0 \frac{dS_1 (t)}{dt}
 = - \pi D \left( \frac{1}{\sqrt{2 \pi Dt}} \right)^3
 = - \pi D n^3 (t).
\label{DifferentialEEA}
\end{equation}

The $n^3$ dependence can also be understood more intuitively. The average excursion length of an exciton at time $t$ is $\sqrt{4 D t}$ for a diffusion constant $2D$, and the average time for excitons to reach the average separation $\frac{1}{n_0}$ can be estimated to be $\frac{1}{4 D n_0^2}$. This results in a decay rate of the exciton density to become $n_0 (\frac{1}{4 D n_0^2})^{-1} = 4 D n_0^3$. Such a dependence arises from the fact that particles in one-dimensional system diffuse compactly in comparison to two- or three-dimensional systems where particles rarely go back to their original positions.\cite{Gennes:1982}

Using the $n^3$ dependence of EEA rate, we can derive an analytical expression for excitation power dependence of PL intensity from a simple rate equation taking into account the balance of exciton generation and decay processes. Considering an enhancement of the linear decay rate caused by end quenching, the rate equation for excitonic density becomes
\begin{equation}
\frac{dn}{dt} = g(z_0) -\frac{1}{\tau P_{\mathrm{I}}(z_0)} n - \pi D n^3.
\label{RateEquation0}
\end{equation}
By approximating $n$ by $N/L$, a steady-state rate equation for excitons in a nanotube can be written as
\begin{equation}
g_0 \tau = \frac{N}{f} - \frac{\pi l^2}{L^2} N^3,
\label{RateEquation}
\end{equation}
where $f = \int g(z_0) P_{\mathrm{I}}(z_0) dz_0 / \int g(z_0) dz_0$ is the fraction of excitons that go through intrinsic decay. The solution to Eq.~(\ref{RateEquation}) is given by the cubic formula
\begin{equation}
N = \sqrt[3]{-q^2 + \sqrt{q^2 + p^3}} + \sqrt[3]{-q^2 - \sqrt{q^2 + p^3}},
\label{CubicFormula}
\end{equation}
where $p = L^2 / 3 \pi f l^2$ and $q = g_0 \tau L^2 / 2 \pi l^2$, and we find that this expression can reasonably reproduce the simulation results [Fig.~\ref{Fig6}(c)]. Using $N \propto I$ and $g_0 \propto P$, Eq.~(\ref{CubicFormula}) can also be used as an analytical expression for excitation power dependence of PL intensity.

\section{Conclusion}
We have investigated exciton diffusion and its effects on luminescence properties of carbon nanotubes by performing rigorous PL characterization on hundreds of individual air-suspended nanotubes. End quenching is evaluated by analyzing the length dependence of PL intensity, and exciton diffusion length and PL action cross section have been obtained. We found that there are clear differences of the values among five different chiralities, and (10,9) nanotubes seem to have a very long diffusion length. We have also investigated the interplay between exciton diffusion, end quenching, and EEA using Monte Carlo simulations, and the rates of exciton generation and decay processes have been determined quantitatively. By comparing the simulations with experiments, we can estimate the absorption cross section and PL quantum yield, and these values also show chirality dependence. From the simulation results, we have found that the EEA rate is proportional to the third power of exciton density, which can be used to derive an analytical expression for excitation power dependence of PL intensity.

The chirality dependence of the exciton properties suggests that it is possible to improve the performance of nanotube photonic devices\cite{Imamura:2013, Noury:2014, Miura:2014} by selecting suitable chiralities. For brighter emission, shorter diffusion length and larger PL action cross section are desirable, while long diffusion length is useful for generating short optical pulses as efficient EEA helps rapid exciton decay. It may be worthwhile studying exciton diffusion in (10,9) nanotubes by a different method, where much longer diffusion length is expected. The cubic dependence of EEA rate underscores the importance of rapid multiple-exciton decay processes, which may potentially be used for efficient single-photon generation.\cite{Hogele:2008}

\begin{acknowledgments}
We thank S. Chiashi and S. Maruyama for the use of the electron microscope. Work supported by KAKENHI (23104704, 24340066, 24654084, 26610080), The Canon Foundation, Asahi Glass Foundation, KDDI Foundation, and the Photon Frontier Network Program of MEXT, Japan. The samples were fabricated at the Center for Nano Lithography \& Analysis at The University of Tokyo. A.I. is supported by MERIT and a JSPS Research Fellowship, and M.Y. is supported by ALPS.
\end{acknowledgments}

\appendix

\section{Derivation of the end quenching survival probability}
\label{app:FirstPassage}
We start by reviewing the use of image method in first-passage processes\cite{Redner} to obtain the density profile $n_1 (z,z_0,t)$ at position $z$ and time $t$ for excitons with initial position $z_0$ and an absorbing boundary located at $z = z_{\mathrm{b}} (<z_0)$. To satisfy the boundary condition $n_1 (z_{\mathrm{b}},z_0, t) = 0$, a negative image particle is placed at $z = 2 z_{\mathrm{b}} -z_0$. This results in 
\begin{equation}
n_1 (z,z_0,t) = p(z,z_0,t) - p(z,2 z_{\mathrm{b}} -z_0,t),
\label{n1}
\end{equation}
where $p (z,z_0,t) = \frac{1}{\sqrt{4 \pi D t}} \exp \left\{ - \frac{(z-z_0)^2}{4 D t} \right\}$ is the normal distribution. The exciton survival probability in this case is
\begin{equation}
S_1 (z_0,t)
= \int_{z_{\mathrm{b}}}^{\infty} n_1 (z,z_0,t) dz
= \mathrm{erf} \left( \frac{z_0 - z_{\mathrm{b}}}{{\sqrt{4 D t}}} \right).
\label{S1}
\end{equation}

In the case of two absorbing boundaries located at $z=-L/2$ and $L/2$, exciton density $n_2 (z,z_0,t)$ should satisfy $n_2 (-L/2,z_0,t) = n_2 (L/2,z_0,t) = 0$. To cancel out the exciton density at the right boundary, we first place a negative image particle at $L-z_0$. As we now need to compensate for this image particle at the left boundary, another positive image particle is placed at $-2L+z_0$. Such an iteration results in an infinite series $\sum_{j=1}^{\infty} (-1)^j p(z, (-1)^j (-jL + z_0), t)$. Also considering a similar procedure starting from the other boundary, 
\begin{widetext}
\begin{equation}
n_2 (z,z_0,t)
= p (z,z_0,t) +
\sum_{j=1}^{\infty} (-1)^j
\left\{ p(z, (-1)^j (-jL + z_0), t) + p(z, (-1)^j (jL + z_0), t) \right\},
\label{n2}
\end{equation}
and therefore the survival probability through end quenching is given by
\begin{equation}
S_{\mathrm{E}} (z_0,t) = \int_{-L/2}^{L/2} n_2 (z,z_0,t) dz
= 1 - \sum_{k=0}^{\infty} (-1)^k
\left\{ \mathrm{erfc} \left[ \frac{\frac{L}{2}(1+2k)-z_0}{\sqrt{4 D t}} \right]
+ \mathrm{erfc} \left[ \frac{\frac{L}{2}(1+2k)+z_0}{\sqrt{4 D t}} \right] \right\}.
\label{SE}
\end{equation}
\end{widetext}

The contributions of additional image particles become more important when $\sqrt{D \tau}$ is longer than the nanotube length $L$. In contrast, when $\sqrt{D \tau} \ll L$, the survival probability can be approximated by
\begin{equation}
S_{\mathrm{E}} (z_0,t) \approx \mathrm{erf} \left( \frac{\frac{L}{2}-z_0}{\sqrt{4Dt}} \right)
 \mathrm{erf} \left( \frac{\frac{L}{2}+z_0}{\sqrt{4Dt}} \right),
\label{SEapprox}
\end{equation}
but using this expression for short nanotube lengths will yield a result inconsistent with solutions of diffusion equations.\cite{Anderson:2013}

Instead of calculating the survival probability, we can obtain the steady-state exciton density profile by
\begin{equation}
n (z)
= \int_{t=0}^{\infty} S_{\mathrm{I}}(t) \int_{z_0=-L/2}^{L/2} g(z_0) n_2 (z, z_0, t) dz_0 dt.
\label{nIntegrated}
\end{equation}
In the case of homogeneous excitation $g(z_0 )=G$, 
\begin{equation}
n (z)
= G \tau \left\{ 1 - \frac{\cosh (z/l)}{\cosh (L/2l)} \right\},
\label{nUniExc}
\end{equation}
which is equivalent to the solutions of diffusion equations.\cite{Xie:2012, Crochet:2012}

\section{Photoluminescence intensity under pulsed excitation}
\label{app:PulsedPL}
Using an approximation $n = N/L$, the $n^3$ dependence of EEA discussed in Sec.~\ref{sec:EEA} results in a rate equation
\begin{equation}
\frac{dN(t)}{dt} = -\frac{1}{\tau f} N(t) -\frac{\pi D}{L^2} N^3 (t),
\label{RateEquationPulsed}
\end{equation}
and this is solved under an initial condition $N(0) = N_0$ to obtain results for pulsed excitation. The solution is
\begin{equation}
N(t) = \left\{ -\frac{\pi f l^2}{L^2} + \left( \frac{\pi f l^2}{L^2} + \frac{1}{N_0^2} \right) \exp \left( \frac{2 t}{\tau f} \right) \right\}^{-\frac{1}{2}},
\label{PulsedSolved}
\end{equation}
and PL intensity is given by
\begin{equation}
I \propto \frac{1}{\tau} \int_{t=0}^{\infty} N(t) dt
= \frac{L}{l} \sqrt{\frac{f}{\pi}} \left\{ \frac{\pi}{2}-\tan^{-1} \left( \frac{L}{l N_0 \sqrt{\pi f}} \right) \right\},
\label{IPulsed}
\end{equation}
which asymptotically reaches a constant value at high powers due to rapid initial nonradiative recombination caused by EEA. The behavior of this expression is very similar to the models given in Refs.~\citenum{Murakami:2009prb2} and \citenum{Srivastava:2009}.

\end{document}